# Fault Tolerant Control of Power Systems in presence of Sensor Failure

S. Khosravani, I. Naziri, A. Afshar IEEE member and M. karrari IEEE senior member
Amir Kabir University of technology
{khosravani_s, iman_naziri, aafshar, karrari }@aut.ac.ir

*Abstract*-- **This paper, addresses Fault Tolerant Control (FTC) of Large Power Systems (LPS) subject to sensor failure. Hiding the fault from the controller allows the nominal controller to remain in the loop. We assume specific faults that violate observability of a subsystem, and we cannot rely on these faulty subsystems when estimating states. We use a new method for reconfiguration control of these faults that lead to unobservability of subsystems. The method proposes augmenting a faulty subsystems with another subsystem(s) until a new subsystem is achieved that is observable. Next, finding the best subsystems among available candidates is considered and using structural analysis methods and grammian definition, a complete algorithm is proposed for FTC of LPS. The proposed approach is applied to the IEEE 14-bus test case and interactions are considered in nonlinear form. Simulation results show that the proposed approach works as intended.**

*Index Terms*--**Fault tolerant control, interconnected large power system, time varying observer, sensor failure, system reconfiguration.**

NOMENCLATURE

| | |
|---|---|
| $\delta_i$ | The power angle of the $i^{th}$ generator, in rad |
| $\omega_i$ | The relative speed of the $i^{th}$ generator, in rad/s |
| $\omega_0$ | The synchronous machine speed, in rad/s |
| $D_i$ | The damping factor, in p.u. |
| $H_i$ | The inertia constant, in $s$ |
| $P_{mi}$ | The mechanical input power, in p.u. |
| $P_{ei}, Q_{ei}$ | Active and the reactive electrical power, in p.u. |
| $E_{fi}$ | The equivalent EMF in the excitation coil, in p.u. |
| $E'_{qi}$ | The transient EMF in the quadrature axis, in p.u. |
| $E_{qi}$ | The EMF in the quadrature axis, in p.u. |
| $T'_{doi}$ | The direct-axis transient time constant, in $s$ |
| $x_{di}$ | The direct axis reactance, in p.u. |
| $x'_{di}$ | The direct axis transient reactance, in p.u. |
| $I_{di}, I_{qi}$ | Direct and the quadrature axis current, in p.u. |
| $I_{fi}$ | The excitation current, in p.u. |
| $x_{adi}$ | The mutual reactance between the excitation coil and the stator coil, in p.u. |
| $B_{ij} + jG_{ij}$ | The mutual reactance between the excitation coil and the stator coil, in p.u. |

## I. INTRODUCTION

The growing dimensions and complexity of the present day technological, environment and societal processes is one of the foremost challenges for engineers. Power systems are amongst the most complex systems ever designed. In particular, they are large scale, nonlinear and have substantial uncertainty in their modeling. In a system Sensor, actuator or process failures may strongly alter the system behavior, which may causes from performance degradation to instability and loss of control. Due to high complexity of power systems, a fault in one component can cause numerous sequential events and if it is not isolated quickly and accurately and managed properly, it can lead to major disruption.

To prevent fault induced losses and minimize the potential risks, new control techniques and design approaches need to be developed in order to deal with faulty system while maintaining overall system stability and performance.

A control system that possesses such a capability is o known as a Fault-Tolerant Control System (FTCS). Increasing system's reliability is the main purpose of FTC.

Quality of sensor measurements and observability of system is highly important in control of power systems. If some sensors are missing, the controllers cannot provide the correct control actions for a plant based on faulty input data. As a result, the plant may have to be tripped off from the power system.

FTC is divided in two main sections contains Passive and Active approaches. The control literature contains a large amount of linear control redesign approaches for AFTC which redesign the nominal controller after FDI. In recent years, Reconfiguration Control Systems (RCS) has obviously attracted attention of many researchers.

Robust performance in large-scale systems [1] linear quadratic regulator method [2] , Pseudo inverse Method [3], [4], perfect model following [5], eigenstructure assignment method [6], adaptive control Approaches [7], are among the most important approaches. FTC of permanent magnet synchronous motor (PMSM) is presented in [8]. O. Wallmark et al. in [9] used an extra inverter leg for actuator redundancy. Sensor FTC has been researched for traction in [10]. Sensitivity analysis towards system order reduction is reviewed in [11]. In [12] for identify state variables which are not available using direct measurement, an estimation based method is proposed. On the other hand some researchers consider this problem from a different point of view. Staroswiecki and Blanke consider the problem as a Fault tolerant structural design in [13]. C. Commault et al. and T. Boukhobza et al.. developed structural methods using graph theory in [14], [15]. Many more papers review monitoring aspect of LPS in order to real time operation and scheduling [23]-[27].

However, to the best of author's knowledge, scarce papers consider a FTC of LPS plant as a large scale complex plant in their researches, which must be considered in many realistic systems. shousung, and Weili proposed an active fault tolerant control for decentralized output feedback control systems considering uncertain interconnected subsystems in [1]. They change the controller parameters to make LSS asymptotically stable. Also, Jing-huan Wang proposed an active Fault tolerant control method using LMI to redesign controller parameters, approach to sensor failure in [16].

In this paper a new approach for FTC of large scale power systems is proposed. We propose to augment faulty subsystems to obtain new healthy subsystems. Using this method one can easily use reconfiguration method for FTC of systems. It is assumed that the model parameters of the faulty system are provided on-line by a diagnosis component, and process diagnosis indicates the fault occurrence and also identifies the fault location and magnitudes. We have all information about model of every subsystem plant.

This paper is organized as follows: The LPS model and linear time varying observer design are described in section II. System reconfigurations with two kinds of fault are proposed in section III. The proposed method is applied to a simulated model of a sample power system in section IV. Conclusion and future works are mentioned in section V.

## II. PROBLEM STATEMENT

In this section a multi machine interconnected system is considered for modeling. Firstly, mathematical modeling and formulation is mentioned. In next section for a class of interconnected system, linear time varying observer is designed due to nonlinearity and uncertainties of subsystems.

### A. Interconnected Power Systems' Dynamics

Mathematical model of a LPS consisting of *n* synchronous machines interconnected trough a transmission network is considered. We use third order nonlinear model derived in [17] for each generator. A model for each sub-system with one generator can be written as follows:

Mechanical dynamics:

$$\dot{\delta}_i = \omega_i \quad (1)$$
$$\dot{\omega}_i = -\frac{D_i}{2H_i}\omega_i + \frac{\omega_0}{2H_i}(P_{mi} - P_{ei}) \quad (2)$$

Electrical dynamics:

$$\dot{E}'_{qi} = \frac{1}{T'_{doi}}(E_{fi} - E'_{qi}) \quad (3)$$

Electrical Equations:

$$E_{qi} = E'_{qi} + (x_{di} - x'_{di})I_{di} \quad (4)$$
$$I_{di} = \sum_{j=1}^{n} E'_{qi}\left(G_{ij}sin(\delta_{ij}) - B_{ij}cos(\delta_{ij})\right) \quad (5)$$
$$I_{qi} = \sum_{j=1}^{n} E'_{qi}\left(B_{ij}sin(\delta_{ij}) + G_{ij}cos(\delta_{ij})\right) \quad (6)$$
$$P_{ei} = \sum_{j=1}^{n} E'_{qi}E'_{qj}\left(B_{ij}sin(\delta_{ij}) + G_{ij}cos(\delta_{ij})\right)$$
$$= E'_{qi}I_{qi} \quad (7)$$
$$Q_{ei} = \sum_{j=1}^{n} E'_{qi}E'_{qj}\left(G_{ij}sin(\delta_{ij}) - B_{ij}cos(\delta_{ij})\right)$$
$$= E'_{qi}I_{qi} \quad (8)$$
$$\delta_{ij} = \delta_i - \delta_j$$
$$E_{qi} = x_{adi}I_{fi}$$

One can easily obtain state space realization of above model as below:

$$\dot{X}_i = A_iX_i + B_iu_i + I_i(t,x,u)$$

Where $A_i \in \Re^{n_i \times n_i}$, $B_i \in \Re^{n_i \times m_i}$ and $I_i \in \Re^{n_i \times k_i}$ denotes nonlinear terms of $i^{th}$ subsystem and nonlinear interaction terms between subsystem $'i'$ and the other subsystems. States and inputs for one subsystem are defined as below:

$$X_i = \begin{bmatrix} x_{i1} \\ x_{i2} \\ x_{i3} \end{bmatrix} = \begin{bmatrix} \Delta\delta_i \\ \Delta\omega_i \\ \Delta E'_{qi} \end{bmatrix}, u_i = \begin{bmatrix} \Delta E_{fi} \\ \Delta P_{mi} \end{bmatrix} \quad (9)$$

If excitation control loop is considered, we will choose mechanical torque ($P_{mi}$) as a constant input. So field voltage is assumed as an input to the system, which is accessible and can be perturbed more easily than the other one [18].

### B. Linear Time Varying Observer

We consider following large scale interconnected form which is introduced in [19].

$$\begin{cases} \dot{X}_i = A_iX_i + B_iu_i + I_i(t,x,u,d) \\ y_i = C_iX_i \end{cases} \quad (10)$$

Where $'d'$ includes bounded disturbance and uncertainties. Other terms are defined as below:

$$X_i = \begin{bmatrix} x_{i1} \\ x_{i2} \\ \vdots \\ x_{in} \end{bmatrix}, A_i = \begin{bmatrix} 0 & 1 & \cdots & 0 \\ \vdots & \vdots & \ddots & \vdots \\ 0 & 0 & 0 & 1 \\ 0 & 0 & 0 & 0 \end{bmatrix}$$

$$B_i = \begin{bmatrix} 0 \\ \vdots \\ 0 \\ 1 \end{bmatrix}, I_i = \begin{bmatrix} I_{i1} \\ I_{i2} \\ \vdots \\ I_{in} \end{bmatrix}, C_i = \begin{bmatrix} 1 & 0 & \cdots & 0 \end{bmatrix}$$

It is obvious that the linear part of (10) is observable regard to the Jordan form of matrices A, C. Hence, system (10) will be observable if nonlinear part satisfies following condition.

**Assumption:** there is a positive constant 'C' such that

$$\begin{cases} |I_{i1}(t,x,u,d)| \leq C(|x_{11}| + |x_{21}| + \cdots + |x_{m1}|) \\ |I_{i2}(t,x,u,d)| \leq C(|x_{11}| + |x_{21}| + \cdots + |x_{m1}| + |x_{m2}|) \\ \vdots \\ |I_{in}(t,x,u,d)| \leq C(|x_{11}| + \cdots + |x_{1n}| + \cdots + |x_{m1}| + \cdots |x_{mn}|) \end{cases} \quad (11)$$

For system (10) under above assumption, linear time varying observer is designed as follows:

$$\begin{aligned} \dot{\hat{x}}_{i1} &= \hat{x}_{i2} + L(t)(e_{i1}) \\ \dot{\hat{x}}_{i2} &= \hat{x}_{i3} + L'(t)(e_{i1}) \\ &\vdots \\ \dot{\hat{x}}_{i(n-1)} &= \hat{x}_{in} + L^{(n-1)}(t)(e_{i1}) \\ \dot{\hat{x}}_{in} &= u_i + L^n(t)(e_{i1}) \end{aligned} \quad (12)$$

Where L(t) is an observer gain parameter to be determined in [20]. A suitable choice for observer gain which be considered in [21] is:

$$\dot{L} = \frac{1}{L^2}(x_{i1} - \hat{x}_{i1})^2, L(0) = 1 \quad (13)$$

And error term is defined as below:

$$e_{ij} = x_{ij} - \hat{x}_{i1}$$

Convergence proof of above observer is in [20] and here it is omitted for brevity.

III. System Reconfiguration Algorithm

This section is divided into two main category based on the observability condition of the faulty subsystem.

*A. Observability of subsystem preserved.*

In this stage if a fault occurs in one of the subsystems and the observability preserved after fault occurrence one can use a simple method [22] to reconstruct healthy sensor information.

The faulty subsystem of the LSS after one fault occurs is presented as below:

$$S_{if}: \begin{cases} \dot{x}_{if} = A_i x_{if} + B_i u_{if} + \sum_{j=1}^{r} h_{ijf}(x) \\ y_{if} = C_{if} x_{if} \end{cases} \quad (14)$$

Through observer design, one can obtain virtual sensor for the LSS based on observed data as below:

$$\tilde{y}_i = P_i y_{if} + (C_i - P_i C_{if})\hat{x}_i$$

Which $\tilde{y}_i$ is reconfigured output sensor data. And $\hat{x}_f(0) = \hat{x}_{f,0}$ and $P_i$ is free parameters to be chosen.

One possible choice of $P_i$ can be an identity matrix with every row corresponding to a faulty sensor is changed to zero. Whenever a fault in sensor occurs, after sensor FDI, switch off the faulty sensor, then switch to related observer and by using a virtual sensor, we have a fault tolerant LSS by reconstructing necessary data. (Require that observability condition preserved). System reconfiguration scheme in this case is shown in Fig. 1.

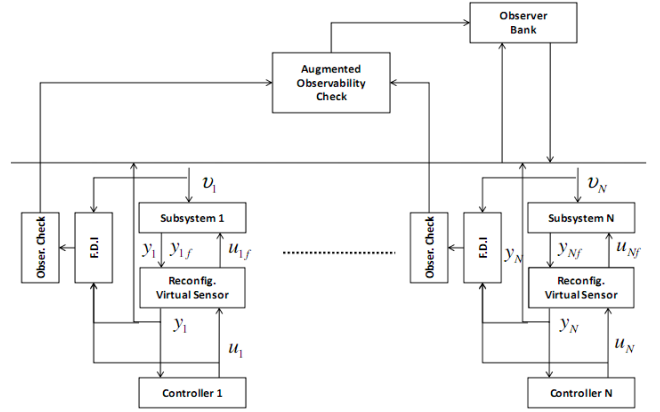

Fig. 1. Reconfiguration Control scheme when Fault occurs in one or more subsystems and Faulty Subsystems is still observable and we can use Virtual Sensors separately in each subsystem.

*B. Observability of subsystem violated*

In this case one may encounter some faults that cause that Observability of subsystems violated. Because of the interconnected nature of considered power plant, one possible solution will be included in a way that other healthy subsystems help the faulty one. Assume one LPS plant contains $N$ subsystems. We propose to use two new blocks.
The first one, for checking observability matrix to ensure that faulty subsystem remains observable or not, under fault condition. We propose to augment subsystems in order to obtain new observable subsystems. Due to interactions which exist between subsystems observability of faulty subsystem can obtain. When all sensors in a subsystem totally fail, the output connectibility of its states is violated. Considering the fact that we are faced with a large scale system, due to large number of available subsystems, checking classical observability condition for all healthy candidate subsystems is very difficult. So we propose to detect structurally unobservable cases by using structural analysis and limit our search collection very effectively.

**Lemma 1:**
Consider the pair of $(A, C)$ of the form

$$A = \begin{bmatrix} A_{11} & A_{12} \\ A_{21} & A_{22} \end{bmatrix}_{n \times n}, C = [C_1 \; \vdots \; 0]$$

Which $A_{11} \in \Re^{k \times k}, A_{12} \in \Re^{(n-k) \times k}, A_{21} \in \Re^{k \times (n-k)}, A_{22} \in \Re^{(n-k) \times (n-k)}$. Then it is easy to see that if $A_{12} = 0$, then $rank\ [C^T\ \ C^T A^T\ \ ...\ \ C^T A^{n-1\ T}]^T$ is less than $n$ (independently of the parameter value of $A_{11}, A_{12}, A_{21}, A_{22}, C_1$).thus the pair is structurally unobservable.

From above lemma, it is inferred directly that, almost for every value of $A_{21}$, for checking non-observability of new augmented subsystems, it is sufficient to consider relation between $A_{11}$ and $A_{22}$ through interaction term $A_{12}$.

Due to special power plant modelling considered in this paper, If there exist a healthy subsystem contains interactions between first state of the faulty subsystem and its states, augmenting these 2 subsystem will be observable. So, in this special case, for checking unobservability of augmented subsystem, it is sufficient to check just first row of interaction matrix $A_{12}$.

Consider the autonomous augmented interconnected system contains subsystem 1 and subsystem 2:

$$\begin{cases} \dot{x}_{New} = \begin{bmatrix} A_{11} & A_{12} \\ 0 & A_{22} \end{bmatrix} x_{NEW} \\ y_{NEW} = [C_1\ \vdots\ 0] x_{NEW} \end{cases}$$

Let the eigenvalues of $A_{11}$ be $\lambda^i_{A_{11}}; i = 1,2, ..., n_{A_{11}}$ and the eigenvalues of $A_{22}$ be $\lambda^j_{A_{22}}; j = 1,2, ..., n_{A_{22}}$.

**Theorem 1:**
The system describe in () is completely observable iff :
i. $(A_{11}, C_1)$ is observable.
ii. $A_{11}$ and $A_{22}$ have distinct eigenvalues, if repeated, then the repeated eigenvalue must have a simple degeneracy, $q = 1$ associated with it, and the following condition must be satisfied
$$rank\ \left(\lambda I - \begin{bmatrix} A_{11} & A_{12} \\ 0 & A_{22} \end{bmatrix}\right) = n - 1\ for\ all\ \lambda$$
iii. The polynomial matrix
$C1\{adj(sI - A_{22})A_{12}adj(sI - A_{11})\}$
contains no common factor $(sI - \lambda^j_{A_{22}}); j = 1,2, ..., n_{A_{22}}$.

**Lemma 2:**
In the case which $A_{11}$ and $A_{22}$ both have repeated eigenvalue $\lambda$ with associated degeneracy $q$, then the system () is completely observable iff the matrix $[C_1^T\ \vdots\ 0]^T$ has at least $q$ linearly independent columns which are not orthogonal to the eigenvectors associated with $\lambda$.

$$q = n - rank\ \left(\lambda I - \begin{bmatrix} A_{11} & A_{12} \\ 0 & A_{22} \end{bmatrix}\right)$$

*Proof:*
The transfer matrix of the system is determined as:

$$G(s) = [C_1\ \ \ 0] \begin{bmatrix} (sI - A_{11})^{-1} & (sI - 22)^{-1} A_{12}(sI - A_{11})^{-1} \\ 0 & (sI - A_{22})^{-1} \end{bmatrix}$$

$$= \frac{1}{\Delta}[C_1\ \ \ 0] \begin{bmatrix} adj(sI - A_{11})\det(sI - A_{22}) \\ 0 \\ adj(sI - A_{22})A_{12}adj(sI - A_{11}) \\ adj(sI - A_{22})\det(sI - A_{11}) \end{bmatrix}$$

$$= \frac{1}{\prod_{i=1}^{n_{A_{11}}}(sI - \lambda^i_{A_{11}}) \prod_{j=1}^{n_{A_{22}}}(sI - \lambda^j_{A_{22}})}$$

$$\times \begin{bmatrix} C1 \left\{ adj(sI - A_{11}) \prod_{j=1}^{n_{A_{22}}}(sI - \lambda^j_{A_{22}}) \right\}, \\ C1\{adj(sI - A_{22})A_{12}adj(sI - A_{11})\} \end{bmatrix}$$

$\Delta = det(sI - A_{22})\det(sI - A_{11})$

For the system to be completely observable, no zero-pole cancellation should be happen. Regards to the observability of pair $(A_{11}, C_1)$, no common factor zero $(sI - \lambda^i_{A_{11}})$ (and $C1 \times adj(sI - A_{11})$) will be presented in transfer function $G(s)$. on the other hand there shall not exist any common factor of $(sI - \lambda^j_{A_{22}})$ in $C1\{adj(sI - A_{22})A_{12}adj(sI - A_{11})\}$. For repeated eigenvalues, first assume that one of subsystems has repeated eigenvalues with multiplicity of $k \leq n_{A_{11}}, k \leq n_{A_{22}}$. and

$$rank\ \left(\lambda_1 I - \begin{bmatrix} A_{11} & A_{12} \\ 0 & A_{22} \end{bmatrix}\right) = n_{A_{11}} + n_{A_{22}} - k.$$

This means that there are $q = 2$ linearly independent eigenvectors. Thus the system () can be transformed to Jordan form as below:

$$\begin{cases} \dot{\mu} = J\mu \\ y = \gamma\mu \end{cases}\ where\ \ J = T_J^{-1} \begin{bmatrix} A_{11} & A_{12} \\ 0 & A_{22} \end{bmatrix} T_J$$
$$\gamma = C \times T_J$$

It shows that first $k$ columns of $(\lambda_1 I - J)$ have zero elements, and therefore $rank(\lambda_1 I - J) = n - k$. to satisfy observability condition following rank condition must hold:

$$rank\ [C(\lambda_1 I - J)] = n$$

It can be obtained if output matrix has at least $k$ linearly independent rows which are not orthogonal to the related eigenvectors. Now, if $\lambda_1$ has a simple degeneracy, it is obvious that, the rank condition in the theorem must satisfy to have an observable augmented subsystem. System reconfiguration scheme in this case is shown in Fig. 2.

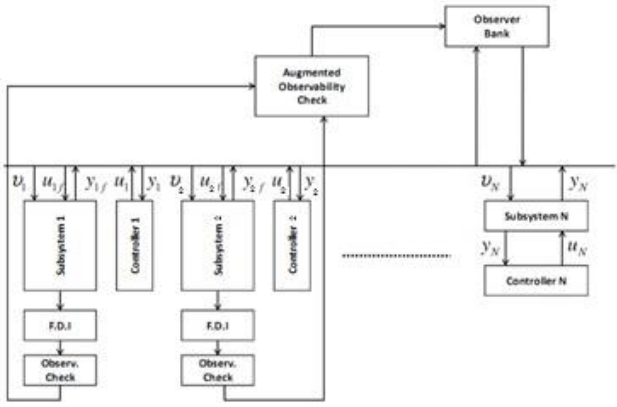

Fig. 2. Reconfiguration Control scheme when Fault occurs and subsystem 1 is not observable anymore, subsystem 1 augmented with subsystem 2 to obtain a new subsystem that is observable

We summarize our proposed method in the following algorithm.

| Algorithm RFTC |  |
|---|---|
| 1 | If the observability condition satisfies use former method that explained using virtual sensor. / stop. |
| 2 | Use Linearized interaction matrix $(A_{ij})$ to augment subsystems. |
| 3 | Check structure of augmented subsystem in order to find and omit unobservable pairs. |
| 4 | Check all of possible two pair of the subsystems contains faulty subsystem, wether if they augmented together they will be observable or not / if it is go to step $n+3$. |
| 5 | check triple subsystems that contain faulty subsystem if that satisfy Observability condition if it is observable go to step $n+3$. |
| ... | ... |
| $n+1$: | Check observability of the system contains $N$ subsystem that it is the whole large scale plant / If it is observable go to step $n+3$. |
| $n+2$: | We cannot use virtual sensor, because system is totally unobservable./ stop. |
| $n+3$: | now system contains $N-\alpha+2$ subsystems which $\alpha$ is the step number of the algorithm. we have a new subsystem such that : $SS_{NEW} : SS_1, SS_2, \cdots, SS_n \quad 2 \leq n \leq N$ New subsystem is observable. Then use the second block for making the new observer for $SS_{NEW}$./ stop. |

Where new subsystem is as below :

$$A_{NEW} = \begin{pmatrix} A_{i1} & A_{i12} & \cdots & A_{i1m} \\ A_{i21} & A_{i2} & & \vdots \\ \vdots & & \ddots & \vdots \\ A_{im1} & A_{im2} & \cdots & A_{imm} \end{pmatrix},$$

$$B_{NEW} = \begin{pmatrix} B_{i1} & 0 & \cdots & 0 \\ 0 & B_{i2} & & \vdots \\ \vdots & & \ddots & \vdots \\ 0 & 0 & \cdots & B_{im} \end{pmatrix},$$

$$C_{NEW} = \begin{pmatrix} 0 & 0 & \cdots & 0 \\ 0 & C_{i2} & & \vdots \\ \vdots & & \ddots & \vdots \\ 0 & 0 & \cdots & C_{im} \end{pmatrix}$$

Where
$A_{NEW} : (n_{i1} + n_{i2} + \cdots + n_{im}) \times (n_{i1} + n_{i2} + \cdots + n_{im})$
$B_{NEW} : (n_{i1} + n_{i2} + \cdots + n_{im}) \times (q_{i1} + q_{i2} + \cdots + q_{im})$
$C_{NEW} : (p_{i1} + p_{i2} + \cdots + p_{im}) \times (n_{i1} + n_{i2} + \cdots + n_{im})$
are state space realizations of new subsystem.

We should linearize nonlinear interaction terms between two subsystems which must be augmented. For linear Sub system's interconnected space state model is derived by linearization (5-8), in the vicinity of an operating point:

$$\dot{x}_i = A_i x_i + B_i u_i + \sum_{j=1}^{n} I_{ij} x_j \qquad (15)$$

After Linearization State matrix will be derived:

$$A_i = \begin{bmatrix} 0 & 1 & 0 \\ a_{21i} & a_{22i} & a_{23i} \\ a_{31i} & 0 & a_{33i} \end{bmatrix}$$

Where

$$a_{21i} = \frac{1}{2H_i} \sum_{j=1}^{n} E'^{0}_{qi} E'^{0}_{qj} \left(G_{ij} sin(\delta_{ij}^{0}) - B_{ij} cos(\delta_{ij}^{0})\right)$$

$$a_{22i} = -\frac{D_i}{2H_i}$$

$$a_{23i} = -\frac{G_{ii} E'^{0}_{qi}}{H_i} - \frac{1}{J_i} \sum_{j=1}^{n} E'^{0}_{qj} \left(B_{ij} sin(\delta_{ij}^{0}) + G_{ij} cos(\delta_{ij}^{0})\right)$$

$$a_{31i} = -\frac{(x_{di} - x'_{di})}{T'_{di}} \sum_{j=1}^{n} E'^{0}_{qj} \left(B_{ij} sin(\delta_{ij}^{0}) + G_{ij} cos(\delta_{ij}^{0})\right)$$

$$a_{33i} = -\frac{1}{T'_{doi}} + \frac{(x_{di} - x'_{di})}{T'_{doi}} B_{ii}$$

And linearized interaction state matrix will be drawn from (5-8):

$$G_{ij} = \begin{bmatrix} 0 & 0 & 0 \\ g_{21i} & 0 & g_{23i} \\ g_{31i} & 0 & g_{33i} \end{bmatrix}$$

Which interaction parameters are defined as follows:

$$g_{21i} = -\frac{1}{J_i} E'^{0}_{qi} E'^{0}_{qj} \left(G_{ij}\sin(\delta_{ij}^{0}) - B_{ij}\cos(\delta_{ij}^{0})\right)$$

$$g_{23i} = -\frac{1}{J_i} E'^{0}_{qi} \left(B_{ij}\sin(\delta_{ij}^{0}) + G_{ij}\cos(\delta_{ij}^{0})\right)$$

$$g_{31i} = -\frac{(x_{di} - x'_{di})}{T'_{di}} E'^{0}_{qj} \left(B_{ij}\sin(\delta_{ij}^{0}) + G_{ij}\cos(\delta_{ij}^{0})\right)$$

$$g_{33i} = -\frac{(x_{di} - x'_{di})}{T'_{di}} \left(G_{ij}\sin(\delta_{ij}^{0}) - B_{ij}\cos(\delta_{ij}^{0})\right)$$

**Remark 1.**
Linearized form of augmented subsystems can easily describe in form (10).

**Remark 2.**
The second block shall contain a bank of all possible observers before using algorithm. Because these procedure is done in offline mode, the assumption of considering an observer bank is not restrictive any the less. Without loss of generality, we can also have a lookup table that contains all type of augmenting subsystems in offline mode, then we know which subsystems could help the others when they need to augment together.

**Remark 3.**
In the proposed algorithm the first block may find more than one subsystem in some cases which have the ability of making new subsystem observable. In this case, we must define a cost function that minimizes or maximize special function including some desired parameters. we need a quality measurement for recovery or an index for degree of observability of new subsystem. We must always consider that one of the best choices is strongly related to amount of energy that we need to transfer deviated state to the nominal trajectory of healthy system. Gramian matrices contain valuable information about energy of observing or controlling for every system. We propose an index based on observability gramian of the subsystem:

$$W_{o-Augmented} = \int_{0}^{\infty} e^{A_{NEW}^T t} C_{NEW}^T C_{NEW} e^{A_{NEW} t} dt \quad (16)$$

Where $W_{co}$ is solution of:

$$W_{o-Aug} A_{NEW} + A_{NEW}^T W_{o-Aug} + C_{NEW}^T C_{NEW} = 0 \quad (17)$$

It holds, the fact that system matrix $A$ is asymptotically stable. We also want to find a new augmented subsystem which has least difference of energy between healthy and faulty condition. One interesting property of the gramian is relation with the plant 2-norm.

$$\|G(s)\|_{h_2}^2 = \frac{1}{2\pi} \int_{0}^{2\pi} trace\left[G(e^{i\theta})\right]^{*}\left[G(e^{i\theta})\right] d\theta$$
$$= trace(CW_c C^T) = trace(B^T W_o B)$$

We propose to define and use cost function as below:

$$J = \alpha \|G(s)\|_{HF}^2 + \xi(\rho_{i\ (1:im)})^2 \quad (18)$$

Where

$$\rho_{i\ (1:im)} = \frac{1}{trace(W_{o-Augmented})} \quad (19)$$

$$N = n_{i1} + n_{i2} + \cdots + n_{im}$$

$$\|G(s)\|_{HF}^2 \triangleq \|G(s)\|_{2H}^2 - \|G(s)\|_{2F}^2$$
$$= trace(B^T(W_{oH} - W_{oF})B)$$
$$= trace(CW_c C^T) - trace(C_F W_c C_F^T)$$

And $\alpha, \xi$ are arbitrary weight coefficients. $k$ is number of subsystems that must be augmented in a special case and $N_i$ is the number of nonzero interaction that subsystem $i$ have when augmented with the other $k$ subsystems.

We also should consider a threshold value that defines the worst case admissible amount of $J$, which even if we find, augmented subsystems that satisfy the observability condition, we could not recover overall system.

## IV. SIMULATION RESULTS

In this section, in order to demonstrate the effectiveness of the proposed method, the Fault tolerant control strategy described above, is applied to large power plant. Firstly, we suppose some kind of fault which when occurs, the Observability of subsystem preserved. Secondly, consider a fault in one of the subsystem occurs that causes totally unobservability of the subsystem. IEEE 14 bus test system is considered to demonstrate effectiveness of the proposed method. We also assume that there exist a FDI block for fault detection and isolation. Our LPS test case contains 5 subsystems as shown in Fig. 3.

The main system is observable and also every 5 subsystems are observable. We assume that in $t = 150$ first fault occurs. Faulty gain is effected on sensor of generator 5, leads to obtain incorrect information about deviation of rotor angle. in this case subsystem 5 remains observable, and exactly after FDI, subsystem switch off sensor(s) and use virtual sensor as defined in first part of section III. We assume that it takes 3 seconds for fault diagnosis and fault isolation.

After $t = 250$ subsystem 5 is totally unobservable. One can easily ignore subsystem 3 from a structural point of view. As mentioned in second part of section III, there are not adequate direct interaction between generator 3 and generator 5 leading to observability of faulty subsystem. And we eliminate subsystem 4 from candidates due to instability of augmented subsystem 5,4. However subsystem 5 affected by these subsystems indirectly. Thus subsystem 1,2 have potential to help faulty subsystem. Checking the cost function results in choosing subsystem 2 to be augmented. Obviously, the assumption of a perfect diagnosis is not realistic so we assume that it takes 1 second for fault diagnosis and fault isolation. According to the former

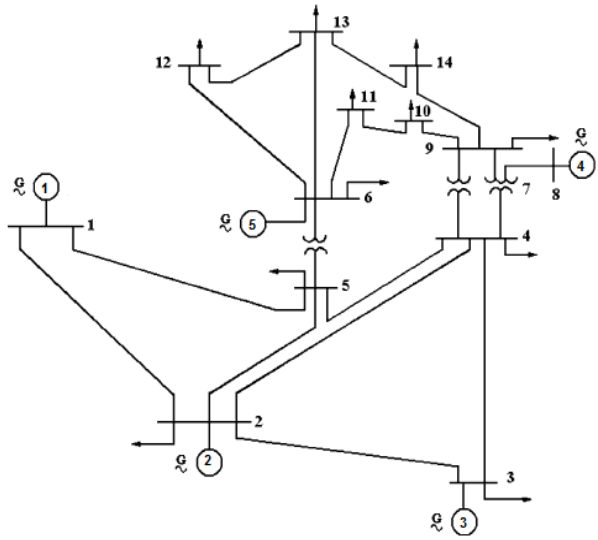

Fig. 3. IEEE 14-bus test system.

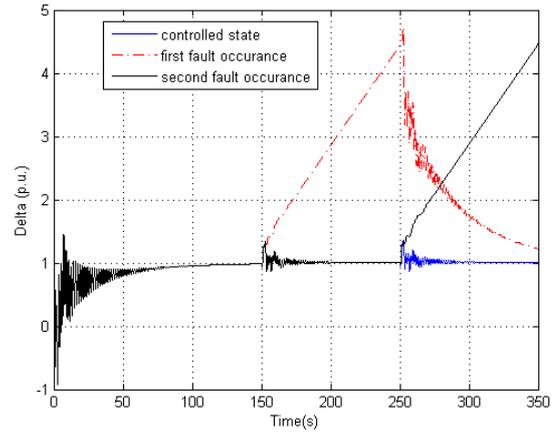

Fig. 4. First state of subsystem 5 with three conditions.

section, we can define cost function as follow:

$$J_{51} = \alpha \left(\frac{1}{3.1151}\right)^2 + \xi(0.1802)^2$$
$$J_{52} = \alpha \left(\frac{1}{6.4649}\right)^2 + \xi(0.2182)^2$$

We consider coefficient factors as below:
$\alpha = 100, \xi = 50$
Then we have
$J_{51} = 11.9290, J_{52} = 4.7732$

We conclude that, in the simulation we must augment subsystem 5 with subsystem 2 to have minimum cost function value.

As the former case we have a gain fault in $t = 150$ but after that in $t = 250$ the second fault occurs.

Fig. 4 and Fig. 5 show the states of subsystem 5 while two types of fault occur. Three conditions are compared in Fig. 4 for the first state. As figure shows, in the first condition system reconfiguration tolerates faults. Second condition happens when a fault occurs in a subsystem while its observability is preserved which is mentioned as the first fault. And the third condition happens when a fault occurs in a subsystem while observability of it is violated.

States of subsystem 5 when two kinds of fault occur in the subsystem are shown in Fig. 4, 5. For first state, three conditions compared in Fig. 4. First condition is when reconfiguration of system tolerates against faults. Second condition is when a fault occurs in a subsystem but observability of the subsystem is preserved (which be mentioned as first fault). And the other condition is when a fault occurs in a subsystem and observability of that subsystem is violated. In second condition, after $t = 250$ reconfiguration method is applied and observability of system is preserved.

Two other states of subsystem 5 are shown in Fig. 5.

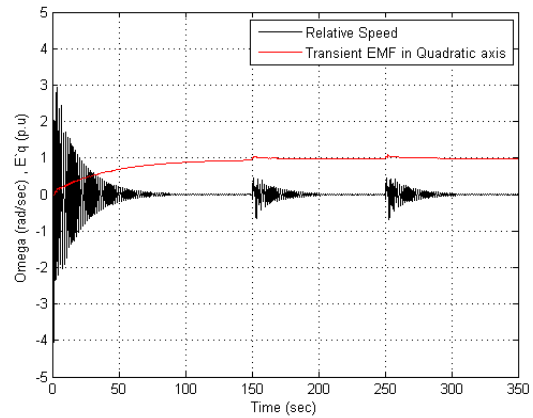

Fig. 5. Second and third states of subsystem 5 when proposed method is applied.

Now we consider another subsystem in which fault has not yet occurred. We mention three conditions for subsystem 3 as well as subsystem 5. Comparisons of different conditions are shown in Fig. 6. Fault occurrence in a subsystem effects obviously on other subsystems performance.

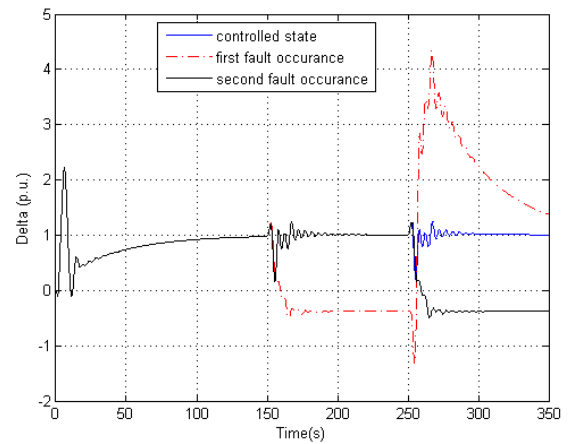

Fig. 6. First state of subsystem 3 with three conditions.

Two other states of subsystem 3 are shown in Fig. 7.

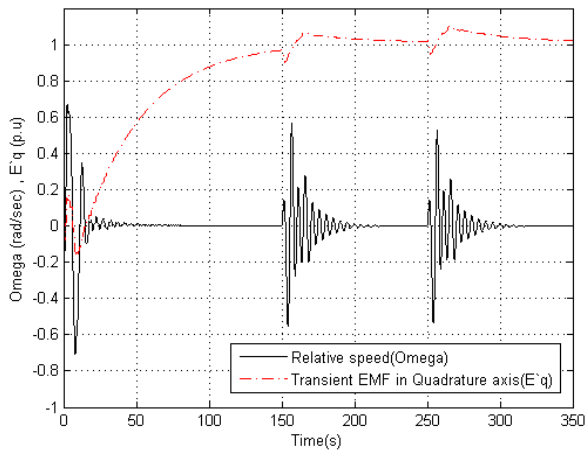

Fig. 7. Second and third states of subsystem 3.

Error convergence of the observer which introduced in second part of section II is shown in Fig. 8. It is seen that the observer is successful in estimating the states.

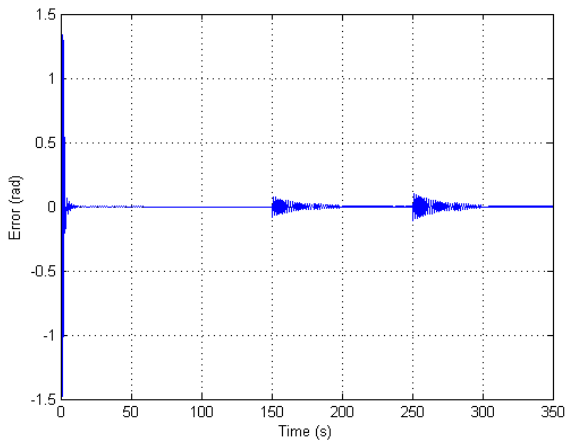

Fig. 8. Error convergence of the introduced observer in subsystem 5.

Simulation result shows that by using the method after a transient, system can continue its nominal behaviour and preserves its stability and performance.

## V. Conclusions

In this paper, a new approach to fault tolerant control of large power systems subject to sensor failure was presented. We propose a method to merge subsystems together when estimation of states is not possible. Although we usually prefer to decouple a LPS to smaller subsystems in order to control them easier. One shall consider that proposed approach only works in faulty condition, and immediately after repairing faulty sensor, system return to its nominal situation, so it is reasonable to augment subsystems and prevent shutting down completely. Simulation results shows that the proposed approach work properly and reconfiguration was done exactly. However there are still many potential aspects for research in this area such as finding a good algorithm to select the best subsystem to be augmented.Indeed, we must mention that there is no guarantee on the system stability during process of replacing faulty sensor with proposed observer.